\documentclass{Interspeech}

\interspeechcameraready 

\usepackage{cite}
\usepackage{amsmath,amssymb,amsfonts}
\usepackage{bm}
\usepackage{graphicx}
\usepackage{textcomp}
\usepackage{xcolor}
\usepackage{dsfont}

\usepackage[nolist]{acronym}
\usepackage{caption}
\usepackage{subcaption}
\usepackage{xcolor}
\usepackage{comment}
\usepackage{cite}
\usepackage{soul}

\usepackage{epstopdf}
\usepackage{xfrac}
\usepackage{pifont}

\usepackage{csquotes}
\usepackage{booktabs}
\usepackage{multirow}
\usepackage{tabularx,ragged2e}
\newcolumntype{Y}{>{\centering\arraybackslash}X}
\usepackage{footmisc}
\usepackage{hyperref}
\usepackage{cleveref}

\usepackage[skins]{tcolorbox}
\usepackage{adjustbox}

\usepackage{array}
\usepackage{empheq}
\usepackage{makecell}
\usepackage{algorithm}
\usepackage{algpseudocode}

\newcommand{\x}[0]{\mathbf{x}}

\title{Non-intrusive Speech Quality Assessment with Diffusion Models \\Trained on Clean Speech}

\author{Danilo}{de Oliveira}
\author{Julius}{Richter}
\author{Jean-Marie}{Lemercier}
\author{Simon}{Welker}
\author{Timo}{Gerkmann}

\affiliation{Signal Processing Group}{University of Hamburg}{Germany}

\email{\{danilo.oliveira, julius.richter, jean-marie.lemercier, simon.welker, timo.gerkmann\}@uni-hamburg.de}

\begin{acronym}
\acro{sgm}[SGM]{score-based generative model}
\acro{snr}[SNR]{signal-to-noise ratio}
\acro{gan}[GAN]{generative adversarial network}
\acro{vae}[VAE]{variational autoencoder}
\acro{ddpm}[DDPM]{denoising diffusion probabilistic model}
\acro{stft}[STFT]{short-time Fourier transform}
\acro{istft}[iSTFT]{inverse short-time Fourier transform}
\acro{sde}[SDE]{stochastic differential equation}
\acro{ode}[ODE]{ordinary differential equation}
\acro{ou}[OU]{Ornstein-Uhlenbeck}
\acro{ve}[VE]{Variance Exploding}
\acro{dnn}[DNN]{deep neural network}
\acro{pesq}[PESQ]{Perceptual Evaluation of Speech Quality}
\acro{se}[SE]{speech enhancement}
\acro{tf}[T-F]{time-frequency}
\acro{elbo}[ELBO]{evidence lower bound}
\acro{rir}[RIR]{room impulse response}
\acro{snr}[SNR]{signal-to-noise ratio}
\acro{lstm}[LSTM]{long short-term memory}
\acro{polqa}[POLQA]{Perceptual Objective Listening Quality Analysis}
\acro{sdr}[SDR]{signal-to-distortion ratio}
\acro{lsd}[LSD]{log-spectral distance}
\acro{sisdr}[SI-SDR]{scale invariant signal-to-distortion ratio}
\acro{estoi}[ESTOI]{Extended Short-Term Objective Intelligibility}
\acro{tcn}[TCN]{temporal convolutional network}
\acro{drr}[DRR]{direct-to-reverberant ratio}
\acro{nfe}[NFE]{number of function evaluations}
\acro{rtf}[RTF]{real-time factor}
\acro{mos}[MOS]{mean opinion scores}
\acro{ema}[EMA]{exponential moving average}
\acro{cnf}[CNF]{continuous normalizing flow}
\acro{visqol}[ViSQOL]{Virtual Speech Quality Objective Listener}
\acro{SRCC}{Spearman's rank correlation coefficient}
\acro{PCC}{Pearson correlation coefficient}
\acro{VB-DMD}{VoiceBank-DEMAND}
\end{acronym}

\begin{document}

\maketitle

\begin{abstract} 
Diffusion models have found great success in generating high quality, natural samples of speech, but their potential for density estimation for speech has so far remained largely unexplored. In this work, we leverage an unconditional diffusion model trained only on clean speech for the assessment of speech quality. We show that the quality of a speech utterance can be assessed by estimating the likelihood of a corresponding sample in the terminating Gaussian distribution, obtained via a deterministic noising process.
The resulting method is purely unsupervised, trained only on clean speech, and therefore does not rely on annotations. Our diffusion-based approach leverages clean speech priors to assess quality based on how the input relates to the learned distribution of clean data. Our proposed log-likelihoods show promising results, correlating well with intrusive speech quality metrics and showing the best correlation with human scores in a listening experiment.
\end{abstract}

\section{Introduction}
\label{sec:intro}

Speech quality estimation is paramount for evaluating algorithms that tackle speech processing tasks, such as speech enhancement, coding, and synthesis.
The golden standard for speech quality estimation is widely considered as the \ac{mos} obtained during listening experiments, where participants are asked to rate audio samples.
However, listening experiments are expensive, time-consuming and can suffer from listener bias if the instructions are not adequately designed.
For these reasons, many instrumental metrics have been proposed to attempt to mimic the result of such listening experiments. 

Instrumental metrics can be handcrafted, which include signal-based metrics such as \ac{sisdr}\cite{roux2019sdr} or \ac{snr}, as well as perceptual metrics integrating some modeling of the human auditory model, like \ac{pesq}~\cite{Rix2001PESQ}, its successor \ac{polqa}\cite{beerends2013perceptual} or \ac{visqol}\cite{chinen2020visqolv3opensource}. Typically, these metrics are intrusive, i.e., they require a reference clean speech signal matching the test utterance. Recording such a clean reference is, however, impractical in real-life scenarios.

In order to avoid relying on reference speech at inference, learning-based metrics based on \acp{dnn} have been proposed, requiring reference speech only at training. These methods typically try to predict the \ac{mos} provided in large labeled speech datasets. These include for example DNSMOS~\cite{reddy2022dnsmosp835}, NISQA\cite{mittag2021nisqa} and NORESQA-MOS~\cite{Manocha2022SpeechQA}. However, these supervised methods might struggle to predict the quality of speech for samples that contain characteristics not encountered during training. Furthermore, supervised methods require large labeled datasets, which can be either inaccessible to the speech research community or susceptible to low-quality annotations.

In this paper, we focus on predicting speech quality in an unsupervised fashion, i.e., training our method only on clean speech data. 
Similar works include SpeechLMScore~\cite{maiti2023speechlmscore} and VQScore~\cite{fu2024selfsupervised}.
SpeechLMScore leverages a language model trained on clean speech tokens and computes the likelihood of the test speech sequence according to the language model vocabulary~\cite{maiti2023speechlmscore}. Lower likelihood will then indicate that the test speech deviates from the clean speech representation of the language model, thereby suggesting low speech quality.
In VQScore~\cite{fu2024selfsupervised}, the authors suggest to train a vector-quantized \ac{vae} on clean speech, and inspect the quantization error at the bottleneck of the model. Since the quantized units define a coarse representation of clean speech, observing a large quantization error suggests that the input speech is not well represented by the discrete codebook and therefore it should be considered of low quality.

In this work, we follow similar ideas as SpeechLMScore and VQScore, but instead choose to use a diffusion model~\cite{ho2020denoising, song2021scorebased} for providing us with a representation of clean speech. 
We compute the likelihood of a test speech by integrating a specific \ac{ode} which, as Song et al. showed~\cite{song2021scorebased}, provides an exact computation of likelihood using a trained diffusion model.
Given that the diffusion model was only trained on clean speech, a test speech utterance of low quality will map to a low-likelihood sample in the terminating Gaussian distribution. 
In a recent work published during the preparation of this manuscript, Emura~\cite{emura2024estimation} showed that the variance of multiple clean speech estimates produced by a diffusion model can be used successfully to estimate the output \ac{sisdr}. 
However, this work only tests the method on clean speech estimates produced by diffusion-based approaches with the same backbone architecture as in the diffusion models that produce the scores.
In comparison to SpeechLMScore~\cite{maiti2023speechlmscore}, our method has no dependence on a choice of speech tokenizer. Rather than using a compressed \ac{vae} latent as in VQScore~\cite{Manocha2022SpeechQA}, we use diffusion models in a less compressed domain, which are more expressive generative models than \acp{vae}. 

We demonstrate that the proposed speech quality estimation correlates well with traditional intrusive metrics such as \ac{polqa}, \ac{sisdr} and \ac{snr}. In particular, our method has a higher correlation to these metrics compared to SpeechLMScore on the traditional VoiceBank-DEMAND noisy speech benchmark~ \cite{valentini2016investigating}. 
Furthermore, we show that, in contrast to SpeechLMScore and VQScore,  our method rates utterances processed by speech enhancement baselines in a similar fashion as intrusive and supervised DNN-based non-intrusive metrics. Code is available online \footnote{\url{https://github.com/sp-uhh/diffusion-sqa}}.

\section{Score-based likelihood estimation}
\label{sec:method}

Score-based generative models~\cite{song2021scorebased} are continuous-time diffusion models relying on stochastic differential equations. 
Such models can learn complex, high-dimensional data distributions such as human speech~\cite{chen2021wavegrad}, natural images~\cite{ho2020denoising, song2021scorebased} or music~\cite{moliner_solving_2022}. 
Score-based models can be considered as iterative Gaussian denoisers. 
At training time, a so-called \textit{forward diffusion process} maps the target data distribution $p(\x)$ to a tractable Gaussian distribution by gradually adding Gaussian-distributed noise to the data $\x \in \mathbb R^n$. 
New data is then generated following the \textit{reverse diffusion process}, which iteratively denoises an initial Gaussian sample until a sample belonging to the target distribution emerges.

Song et al.~\cite[App. D]{song2021scorebased} show that every stochastic diffusion process has a corresponding deterministic process described by an \ac{ode} whose trajectories share the same marginals $p_t(\x)$ as the original process. This specific \ac{ode} is named the \emph{probability flow \ac{ode}}, and it continuously increases (forward in time) or decreases (backward) the level of noise in the data.
Karras et al.~\cite{karras2022elucidating} formulate the probability flow ODE as
\begin{equation}
    \mathrm{d}\x = -\dot\sigma(t) \sigma(t) \nabla_\x \log p_t(\x) \mathrm{d} t \text{,}
    \label{eq:ode}
\end{equation}
where $\sigma(t)$ is a noise schedule defining the level of noise at time $t$ and $\dot\sigma(t)$ is its derivative with respect to $t$. $\nabla_\x \log p_t(\x)$ is the \emph{score function}, a vector field pointing in the direction of higher density of data:
\begin{equation}
    \nabla_\x \log p_t(\x) = \frac{ D_\theta(\x; t) - \x }{ \sigma(t)^2}
\end{equation}
Here, $D_\theta(\x; t)$ is a denoiser function implemented as a neural network $F_\theta(\x; t)$. In order to stabilize and facilitate the training of the model in spite of varying levels of noise, a series of $\sigma$-dependent scaling operations $c_\text{in}$, $c_\text{noise}$, $c_\text{out}$ and $c_\text{skip}$ are used, preconditioning inputs and outputs of the network to have unit variance, as well as a skip connection to avoid amplifying errors in $F_\theta$. When $\sigma(t) = t$ as suggested in Karras et al.~\cite{karras2022elucidating}, \Cref{eq:ode} can be  written as
\begin{equation}
    \mathrm{d}\x = \underbrace{\frac{\x - D_\theta(\x; t)}{t}}_{\bm{f}_\theta(\x; t)}\mathrm{d}t,
    \label{eq:edm_ode}
\end{equation}
with
\begin{equation}
    D_\theta(\x; t) = c_\text{skip}(t)\x + c_\text{out}(t) F_\theta \big( c_\text{in}(t)\x; c_\text{noise}(t) \big).
\end{equation}
In \Cref{eq:edm_ode} we defined the \emph{drift} $\bm{f}_\theta(\x; t)$ that is central to the calculations in \Cref{eq:change_var,eq:logp0,eq:ivp}. Song et al.\cite{song2021scorebased} leverage the probability-flow \ac{ode} associated with their \ac{sde} to compute the log-likelihood of the input data. This is similar to the density estimation procedure from neural \acp{ode}, leveraging the \emph{instantaneous change of variables} formula~\cite{chen2018neural}:
\begin{equation}
    \frac{\partial \log p_t(\x)}{\partial t} = - \text{Tr} \left( \frac{\partial \bm{f_\theta}(\x, t)}{\partial \x} \right)
    \label{eq:change_var}
\end{equation}
Following Grathwohl et al.~\cite{grathwohl2018scalable}, by integrating \Cref{eq:change_var} for the ODE in \Cref{eq:edm_ode}, we get the following expression for the log density of the data:
\begin{equation}
    \log p_0(\x) = \log p_T(\x_T) + \int_0^T \text{Tr} \left( \frac{\partial \bm{f}_\theta(\x_t; t)}{\partial \x_t} \right)\mathrm{d}t
    \label{eq:logp0}
\end{equation}
with initial value $\mathbf x_0 := \mathbf x$ and $\mathbf x_T = \int_0^T  \bm{f}_\theta(\mathbf x_t; t) \mathrm dt$.
 Additionally, the authors show that the trace of the Jacobian matrix $\frac{\partial \bm{f}_\theta}{\partial \x_t}$ can be efficiently computed using the Hutchinson estimator~\cite{hutchinson1990stochastic}:
\begin{equation}
    \text{Tr} \left( \mathbf{A} \right) = \mathbb{E}_{p(\bm{\epsilon)}}[\bm{\bm{\epsilon}}^\top \mathbf{A} \bm{\epsilon}]\text{,}
    \label{eq:hutchinson-trace}
\end{equation}
where the distribution $p(\bm{\epsilon})$ must satisfy $\mathbb{E}[\bm{\epsilon}] = 0$ and $\text{Cov}_{p(\bm{\epsilon})}[\bm{\epsilon}] = \mathbf{I}$. This avoids the computation of a separate derivative for each element of the diagonal of the Jacobian.

The following coupled initial value problem is then solved:
\begin{equation}
    \begin{bmatrix}
        \x_T \\
        \log p_0(\x) - \log p_T(\x_T)
    \end{bmatrix} = \int_0^T \begin{bmatrix}
        \bm{f}_\theta(\x_t; t)\\
        \bm{\epsilon}^\top \frac{\partial \bm{f}_\theta(\x_t; t)}{\partial \x_t}~\bm{\epsilon}
    \end{bmatrix} \text{d}t
    \label{eq:ivp}
\end{equation}
with initial value $\begin{bmatrix}\x_0 & 0 \end{bmatrix}^\top$. Both equations are solved simultaneously using the same solver, as explained in \Cref{alg:loglikelihood}. The vector-Jacobian product $\bm{\epsilon}^\top \frac{\partial \bm{f}_\theta}{\partial \x_t}$ can be evaluated at roughly the same cost as a computation of $\bm{f}_\theta(\x_t; t)$ by performing reverse-mode automatic differentiation, having already computed the forward pass when solving for $\x_t$. $\log p_T$ is obtained from the probability density function of a multivariate Gaussian.

\begin{algorithm}
    \caption{Log-likelihood}
    \label{alg:loglikelihood}
    \begin{algorithmic}[1]
        \State $\x\gets \mathrm{Data}$
        \State $\bm{\epsilon} \sim \mathrm{Rademacher}$
        \State $\Delta\log p \gets 0$
        \State $t \gets 0$
        \While{$t<T$}
            \State $\mathrm{d}\x \gets \bm{f}_\theta(\x;t)$
            \State $\mathrm{d}\Delta\log p \gets \mathrm{Hutchinson(\mathrm{d}\x, \x, \epsilon)}$
            \State $\x \gets \x + \mathrm{d}\x \cdot \mathrm{d}t$ 
            \State $\Delta\log p \gets \Delta\log p + \mathrm{d}\Delta\log p \cdot \mathrm{d}t$
            \State $t \gets t + \Delta t$
        \EndWhile
        \State \textbf{return} $\log p \gets \log p_T + \Delta\log p$
    \end{algorithmic}
\end{algorithm}
\begin{table*}[t]
    \centering
    \caption{Pearson (PCC) and Spearman's rank (SRCC) correlation coefficients between supervised/unsupervised metrics and human-assigned scores on the EARS-WHAM (matched case) and VoiceBank-DEMAND (mismatched) noisy test sets. ($-$)SpeechLMScore indicates that the correlations have been flipped to agree to the other metrics where higher is better.}
    \label{tab:correlations-noisy}
    \begin{tabular}{llcccccccc}
        \toprule
        & & \multicolumn{4}{c}{EARS-WHAM} & \multicolumn{4}{c}{VoiceBank-DEMAND} \\
        \cmidrule(lr){3-6}\cmidrule(lr){7-10}
        & & \multicolumn{2}{c}{POLQA} & \multicolumn{2}{c}{SI-SDR} & \multicolumn{2}{c}{POLQA} & \multicolumn{2}{c}{SI-SDR} \\
        \cmidrule(lr){3-4}\cmidrule(lr){5-6} \cmidrule(lr){7-8}\cmidrule(lr){9-10}
        & Measure & PCC & SRCC & PCC & SRCC & PCC & SRCC & PCC & SRCC \\
        \midrule
        \multirow{2}{*}{Supervised} & NISQA~\cite{mittag2021nisqa} & \textbf{0.797} & \textbf{0.816} & 0.689 & 0.712 & \textbf{0.897} & \textbf{0.896} & \textbf{0.608} & \textbf{0.591} \\
        & DNSMOS OVRL~\cite{reddy2022dnsmosp835} & 0.667 & 0.711 & 0.625 & 0.650 & 0.776 & 0.828 & 0.542 & 0.569 \\
        \midrule
        \multirow{3}{*}{Unsupervised} & VQScore~\cite{fu2024selfsupervised} & 0.723 & 0.755 & \textbf{0.804} & \textbf{0.821} & 0.837 & 0.841 & 0.539 & 0.537 \\
        & ($-$)SpeechLMScore~\cite{maiti2023speechlmscore} & 0.761 & 0.779 & 0.733 & 0.760 & 0.702 & 0.681 & 0.471 & 0.428 \\
        & Diff. Log-likelihood (ours) & 0.640 & 0.667 & 0.617 & 0.633 & 0.831 & 0.835 & 0.489 & 0.498 \\
        \bottomrule
    \end{tabular}
\end{table*}
\begin{table}[h]
    \centering
    \caption{Pearson (PCC) and Spearman's rank (SRCC) correlation coefficients between supervised/unsupervised metrics and human-assigned scores on the WSJ0-CHiME3 listening experiment results of Richter et al.~\cite{richter2023speech}. ($-$)SpeechLMScore indicates that the correlations have been flipped to agree to the other metrics where higher is better.}
    \label{tab:correlations-listening}
    \adjustbox{max width=\linewidth}{
    \begin{tabular}{llcc}
        \toprule
        & & \multicolumn{2}{c}{Listening Scores} \\
        \cmidrule{3-4}
                                    & Measure                      & PCC            & SRCC           \\
        \cmidrule(lr){1-2} \cmidrule(lr){3-4}
        \multirow{2}{*}{Supervised} & NISQA                       & 0.822          & 0.840          \\
                                    & DNSMOS OVRL                 & 0.785          & 0.837          \\
        \cmidrule(lr){1-2} \cmidrule(lr){3-4}
        \multirow{3}{*}{Unsupervised} & ($-$)SpeechLMScore          & 0.762          & 0.818          \\
                                    & VQScore                     & 0.798          & 0.724          \\
                                    & Diff. Log-likelihood (ours) & \textbf{0.902} & \textbf{0.899} \\
        \bottomrule
    \end{tabular}
    }
\end{table}

\section{Implementation details}
The framework used in this work to train the diffusion model is the denoising score matching formulation proposed by Karras et al.~\cite{karras2022elucidating}. Our neural network follows the ADM architecture~\cite{dhariwal2021diffusion}, with the architectural and training improvements subsequently proposed in~\cite{karras2024analyzing}, in particular the \emph{magnitude preserving layers}, which keep the magnitudes of activations in the model controlled during training. We reduce the model size by using only 3 resolutions and employing one residual block per resolution, resulting in a model with 49$\,$M parameters. We set the exponential moving average length to $0.08$ using \emph{post-hoc} reconstruction method from~\cite{karras2024analyzing}. As in \cite{karras2022elucidating}, we employ a second order solver with 32 steps.

For estimation of the trace, we sample $\bm{\epsilon}$ from the Rademacher distribution, which is commonly used with the Hutchinson trace estimator~\cite{hutchinson1990stochastic}; it is a discrete distribution taking two possible values $\{1,-1\}$, each occurring with equal probability, therefore it satisfies the conditions specified in \Cref{sec:method}. The authors of~\cite{grathwohl2018scalable} found the variance of the log-likelihood induced by the trace estimator to be lower than $10^{-4}$ on a validation set, and we confirm this finding on our model by running it with different seeds. We therefore compute the trace using only one noise vector. In order to obtain a consistent and reproducible result, we keep a fixed seed for $\bm\epsilon$. Similarly to Song et al.~\cite{song2021scorebased} and other works showing log-likelihood scores, we normalize the likelihoods by the number of elements in the sample; in this case, time-frequency bins.

We train the model on the EARS dataset~\cite{richter2024ears}, following the train--validation--test split performed for the EARS-WHAM set proposed in the same work~\cite{richter2024ears}. This training set consists of approximately 87 hours of clean, anechoic English speech data with different speaking styles. Here, the audio is downsampled to 16$\,$kHz and transformed into mel spectrograms, with 80 mel bands and Hann windows of length 64$\,$ms with 75\% overlap. The dynamic range of the spectrogram is compressed using the logarithm, and the values are then scaled to match mean 0 and standard deviation of 0.5, using the statistics computed on the training set. 

During training, we sample segments of 4 seconds in length, with random starting indices. The segments are sampled from the dataset and the model is trained on approximately 37$\,$M sampled segments, with batch size 128. We perform evaluations on the EARS-WHAM test set at 16$\,$kHz. The test set contains 6 speakers and input \acp{snr} randomly sampled in a range of $[-2.5, 17.5]\,$dB. We additionally report results on the \ac{VB-DMD} test set~\cite{valentini2016investigating} also downsampled to 16kHz, as a mismatched condition for the model. The test set contains two speakers and noise at 2.5, 7.5, 12.5 and 17.5$\,$dB \ac{snr}. Lastly, we make use of the samples from the listening experiment of Richter et al.~\cite{richter2023speech}, where participants were asked to rate randomly selected samples from the WSJ0-CHiME3 dataset~\cite{richter2023speech}, reconstructed by three different speech enhancement models.

\begin{figure}[h]
    \centering
    \begin{subfigure}{0.85\linewidth}
        \includegraphics[width=\linewidth]{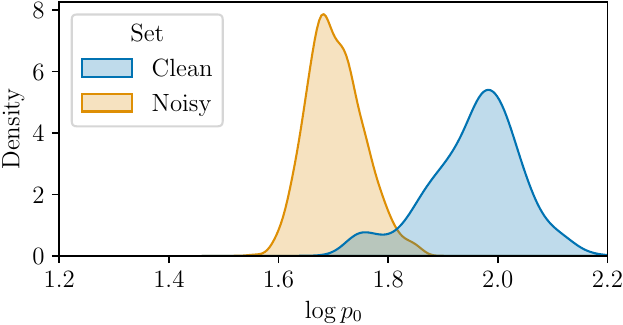}
        \caption{EARS-WHAM}
        \label{fig:hist_earswham}
    \end{subfigure}
    \par\vspace{1em}
    \begin{subfigure}{0.85\linewidth}
        \includegraphics[width=\linewidth]{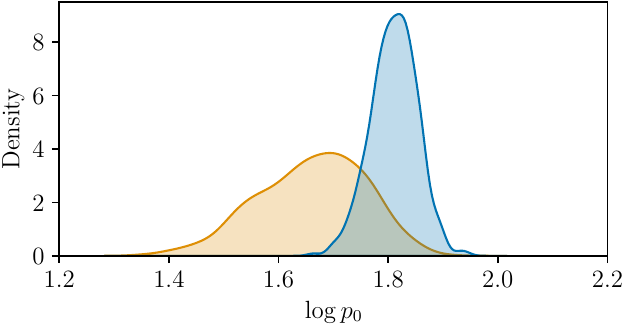}
        \caption{VB-DMD}
        \label{fig:hist_vbdmd}
    \end{subfigure}
    \caption{Histogram of diffusion-based log-likelihood values for noisy and clean test data}
    \label{fig:hist}
\end{figure}
\begin{table*}[ht]
    \centering
    \caption{Evaluation results on the listening experiment samples from Richter et al.~\cite{richter2023speech}, with WSJ0+CHiME3 data. The listening scores are presented in a scale from 0 to 100.}
    \label{tab:abs_listening}
    \adjustbox{max width=\textwidth}{
    \begin{tabular}{lccccccccc}
        \toprule
        & \multicolumn{3}{c}{Intrusive} & \multicolumn{5}{c}{Non-Intrusive} & Listening\\
        \cmidrule(lr){2-4}\cmidrule(lr){5-9}\cmidrule(lr){10-10}
        & POLQA $\uparrow$ & PESQ $\uparrow$ & SI-SDR $\uparrow$ & \makecell{DNSMOS\\OVRL} $\uparrow$ & NISQA $\uparrow$ & SpeechLMScore $\downarrow$ & VQScore $\uparrow$ & \makecell{Diffusion\\Log-likelihood} $\uparrow$ & Score $\uparrow$ \\
        \midrule
        Clean & --- & --- & --- & $3.34\pm0.12$ & $4.24\pm0.27$ & $1.35\pm0.09$ & $0.720\pm0.007$ & $1.90\pm0.04$ & $98.9\pm1.3$\\
        \midrule
        SGMSE+ & $\mathbf{3.32\pm0.35}$ & $2.44\pm0.33$ & $12.5\pm1.9$ & $\mathbf{3.26\pm0.09}$ & $\mathbf{3.82\pm0.29}$ & $\mathbf{1.49\pm0.08}$ & $0.715\pm0.009$ & $\mathbf{1.87\pm0.03}$ & $\mathbf{88.2\pm3.5}$\\
        ConvTasNet & $3.15\pm0.35$ & $2.41\pm0.34$ & $\mathbf{15.4\pm2.1}$ & $3.25\pm0.15$ & $3.70\pm0.32$ & $1.53\pm0.10$ & $\mathbf{0.719\pm0.012}$ & $1.80\pm0.04$ & $75.9\pm7.7$\\
        MetricGAN+ & $2.92\pm0.32$ & $\mathbf{2.61\pm0.22}$ & $6.4\pm2.2$ & $2.83\pm0.25$ & $3.34\pm0.37$ & $1.61\pm0.11$ & $0.681\pm0.023$ & $1.65\pm0.06$ & $49.7\pm9.9$\\
        \midrule
        Noisy & $1.97\pm0.39$ & $1.22\pm0.10$ & $2.52\pm1.73$ & $1.89\pm0.46$ & $1.65\pm0.59$ & $2.11\pm0.31$ & $0.610\pm0.020$ & $1.65\pm0.04$ & $27.2\pm4.0$\\
        \bottomrule
    \end{tabular}}
\end{table*}
\begin{table*}[ht]
    \centering
    \caption{Evaluation results on a predictive method (Demucs) and a generative method (SGMSE+). These correspond to the models reported by the EARS-WHAM benchmark~\cite{richter2024ears}, with the enhanced outputs downsampled from 48kHz to 16kHz.}
    \label{tab:abs_earswham}
    \adjustbox{max width=\textwidth}{
    \begin{tabular}{lccccccccc}
        \toprule
        & \multicolumn{3}{c}{Intrusive} & \multicolumn{5}{c}{Non-intrusive} \\
        \cmidrule(lr){2-4}\cmidrule(lr){5-9}
        & POLQA $\uparrow$ & PESQ $\uparrow$ & SI-SDR $\uparrow$ & \makecell{DNSMOS\\OVRL} $\uparrow$ & NISQA $\uparrow$ & SpeechLMScore $\downarrow$ & VQScore $\uparrow$ & \makecell{Diffusion\\Log-likelihood} $\uparrow$ \\
        \midrule
        Clean & --- & --- & --- & $3.12\pm0.36$ & $4.11\pm0.73$ & $1.38\pm0.18$ & $0.719\pm0.020$ & $1.96\pm0.09$ \\
        \midrule
        SGMSE+~\cite{richter2023speech} & $\mathbf{3.45}\pm0.67$ & $\mathbf{2.53}\pm0.63$ & $16.81\pm4.50$ & $\mathbf{3.13}\pm0.36$ & $\mathbf{4.22}\pm0.74$ & $\mathbf{1.40}\pm0.19$ & $0.723\pm0.020$ & $\mathbf{1.92}\pm0.09$ \\
        Demucs~\cite{rouard2023hybrid} & $3.17\pm0.67$ & $2.40\pm0.59$ & $\mathbf{16.95}\pm4.37$ & $3.08\pm0.36$ & $3.72\pm0.75$ & $1.42\pm0.20$ & $\mathbf{0.728}\pm0.019$ & $1.84\pm0.08$ \\
        \midrule
        Noisy & $1.82\pm0.53$ & $1.25\pm0.22$ & $6.02\pm6.12$ & $2.08\pm0.66$ & $2.02\pm0.70$ & $2.06\pm0.26$ & $0.619\pm0.027$ & $1.70\pm0.05$ \\
        \bottomrule
    \end{tabular}}
\end{table*}

\section{Results and discussion}

In \Cref{tab:correlations-noisy}, we show correlations with intrusive metrics on the noisy data from the EARS-WHAM and \ac{VB-DMD} test sets, compared to the other non-intrusive baselines. We report the \ac{PCC} and \ac{SRCC}, quantifying linear and monotonic relationships, respectively. While the correlation of the diffusion-based log-likelihood is generally below that of the other metrics on EARS-WHAM, it performs similarly to VQScore and better than SpeechLMScore on \ac{VB-DMD}.

\Cref{fig:hist_earswham} shows how the distribution of the log-likelihoods for clean EARS-WHAM test data compare to that of the noisy. We note that the likelihoods are densities, meaning that they are not bounded to a maximum value of 1. Consequently, the logarithm is not necessarily negative. The values for clean data have a large standard deviation, which hints that the clean speech in the test set has variations which are not completely modeled by the neural network. Nevertheless, the two distributions are clearly separated, as one would wish for in a quality metric. For the \ac{VB-DMD} test data, the mean scores of clean speech are lower, which is to be expected, since the data is mismatched from training. Additionally, the distribution is narrower, a finding that can be explained by the less varied expressivity if compared to EARS. Concerning the noisy set, although the data is again mismatched, the larger proportion of higher \acp{snr} mixtures seems to be working in the opposite direction, resulting in a wider distribution.

\Cref{tab:correlations-listening} shows the correlations of listening scores from the experiment of Richter et al.~\cite{richter2023speech} with DNN-based metrics. Supervised methods were trained with MOS labels as targets and non-paired methods were trained in an unsupervised way, with only clean speech data. The numbers show the highest correlation between our proposed approach and the scores given by audio experts, confirming the effectiveness of the method. 

One important aspect to analyze is how a measure handles corruptions from an enhancement model, which can have different behaviors depending on the training paradigm~\cite{lemercier2023analysing, deoliveira2023behavior, pirklbauer2023evaluation}. To paint a complete picture of the models' performances, we evaluate them with a set of non-intrusive and intrusive metrics~\cite{deoliveira2024pesqetarian}, presenting the results on a system level. In the non-intrusive group of metrics, we report VQScore and SpeechLMScore (in perplexity, so lower is better), as well as metrics trained in a supervised fashion, with \ac{mos} labels. In the intrusive subset, we employ PESQ~\cite{Rix2001PESQ} and POLQA~\cite{beerends2013perceptual}, as well as \ac{sisdr}~\cite{roux2019sdr}. \Cref{tab:abs_listening} displays the scores of the listening experiment data, indicating that our method's scores agree with the listeners' by giving a preference to SGMSE+. Almost all non-intrusive metrics point SGMSE+ as the leading model as well, with the exception of VQScore. Additionally, out of the intrusive metrics, only POLQA also reflects this preference.

For an evaluation with more samples, we compare our method with other known objective metrics on a speech enhancement benchmark. In \Cref{tab:abs_earswham} we show such a comparative evaluation on the EARS-WHAM set, comparing the generative method SGMSE+~\cite{richter2023speech, richter2024ears} against the predictive method Demucs~\cite{rouard2023hybrid}, both trained on EARS-WHAM data.
To make it compatible with the metrics, we downsample the enhanced files to 16kHz. Here we can see that, along with DNSMOS and NISQA, log-likelihood favors SGMSE+ over Demucs, whereas VQScore~\cite{fu2024selfsupervised} and SpeechLMScore~\cite{maiti2023speechlmscore} show a only a minor difference between these two evaluated enhancement methods, and even produce values slightly in favor of Demucs. The log-likelihood is therefore the only non-intrusive method trained without access to paired \ac{mos} data that correctly reflects the clear human listener preference for the method SGMSE+ over Demucs reported in~\cite{richter2024ears}. Furthermore, the log-likelihood also shows better alignment with POLQA and PESQ scores.

Even though quality assessment measures generally do not face limited computational constraints, diffusion-based methods such as ours tend to be more computationally demanding than predictive methods. Due to the multiple calls of the score network required to iteratively solve \Cref{eq:ivp}, inference is slower than the baselines. At 32 steps, evaluating 100 utterances takes $\sim$6 minutes, while VQScore takes $\sim$3 seconds. Nevertheless, there is active research in diffusion models that aims to reduce the number of steps without sacrificing performance.

\section{Conclusion}

We proposed a speech quality estimator based on unconditional score-based diffusion models trained on clean speech only. 
Using the natural likelihood computation abilities of score-based models, the proposed estimator can estimate the quality of speech utterances without having any access to paired data.
The resulting measure is non-intrusive and yet correlates well with intrusive metrics on noisy speech benchmarks.
When evaluating utterances processed by speech enhancement baselines, our method showed the highest correlation with the scores assigned by human listeners.

\section{Acknowledgement}

Funded by the Deutsche Forschungsgemeinschaft (DFG, German Research Foundation) – 498394658. We acknowledge the support by the DFG in the transregio project Crossmodal Learning (TRR 169) and DASHH (Data Science in Hamburg -- Helmholtz Graduate School for the Structure of Matter) with Grant-No. HIDSS-0002.
We would like to thank J. Berger and Rohde\&Schwarz SwissQual AG for their support with POLQA.

\bibliographystyle{IEEEtran}
\bibliography{refs}

\end{document}